\documentclass{cernrepNum}   
     
\usepackage{epsfig}   
\newcommand{\doe}{$\delta\,\!^{18}$O}
\newcommand{\degc}{$^\circ$C}
\newcommand{\beten}{$^{10}$Be}
\newcommand{\cft}{$^{14}$C}

\newcommand{\htos}{H$_{2}\,\!^{16}$O}
\newcommand{\htoe}{H$_{2}\,\!^{18}$O}

\begin{document}

\begin{center}
	{\large EUROPEAN ORGANIZATION FOR NUCLEAR RESEARCH} \\[8mm]
\end{center}

\begin{flushright}
{\large CERN--PH--EP/2004--027} \\ 
{\large 18 June 2004}
\end{flushright}

\begin{center}
\textbf{\large THE GLACIAL CYCLES AND COSMIC RAYS}

\vspace{8mm}         

{\large J. Kirkby$^1$, A. Mangini$^2$, R.A. Muller$^3$} \\[2ex]
\textit{\normalsize $^1$CERN, CH-1211, Geneva, Switzerland \\
$^2$Institute for Environmental Physics, University of Heidelberg, Germany D-69120 \\
$^3$Department of Physics, University of California, Berkeley, California 94720, USA}

\vspace{22mm}         
\textbf{\large Abstract}
\end{center}

\begin{center}
\begin{minipage}{140mm} 
The cause of the glacial cycles remains a mystery.  The origin is widely accepted to be astronomical since paleoclimatic archives contain strong spectral components that match the frequencies of Earth's orbital modulation.  Milankovitch insolation theory contains similar frequencies and has become established as the standard model of the glacial cycles.  However, high precision paleoclimatic data have revealed serious discrepancies with the Milankovitch model that fundamentally challenge its validity and re-open the question of what causes the glacial cycles.  We propose here that the ice ages are initially driven not by insolation cycles but by cosmic ray changes, probably through their effect on clouds.  This conclusion is based on a wide range of evidence, including results presented here on speleothem growth in caves in Austria and Oman, and on a record of cosmic ray flux over the past 220~kyr obtained from the \beten\ composition of deep-ocean sediments. 
\end{minipage}
\end{center}
                                           
\vspace{2cm}         

\begin{center}
\textit{\large Submitted to \textit{Earth and Planetary Science Letters}}
\end{center}

\thispagestyle{empty} 


\pagenumbering{roman}  
\setcounter{page}{1}  

\pagestyle{plain}     

\newpage \mbox{} \newpage 

 
\newpage
 
\pagenumbering{arabic}  
\setcounter{page}{1}  

\section{INTRODUCTION}
\label{sec_introduction}

The most important clue for identifying the cause of glacial cycles is the spectral purity of their periodicity, as recorded in \doe, a proxy for global ice volume \cite{do18}.   For the past million years, the glacial pattern is dominated by a precise 100~kyr cycle, and for the million years before that, by an equally precise 41~kyr cycle.  These match the major frequencies of Earth's orbital changes, and so it is generally accepted that the primary driver for past climate change was astronomical.  A linkage between orbit and climate is provided by the Milankovitch model, which states that melting of the northern ice sheets is driven by peaks in Northern Hemisphere summer insolation (solar heating).  The insolation model has gained widespread acceptance since it naturally includes spectral components at the orbital modulation frequencies.

However, as better data have become available, difficulties have arisen with the Milankovitch model.  Insolation, by itself, includes a 41~kyr cycle but no significant 100~kyr variation and so, to enhance it, a nonlinear response of climate to insolation forcing has been postulated \cite{imbrie80,muller00,muller00a}.  An expected 400~kyr cycle is not observed, and so it is simply assumed to be suppressed.  The modulation of the 100~kyr cycles does not follow the expected pattern---a disagreement known as the ``Stage-11 problem'', when a major glacial termination occurred during a period of minor insolation changes around 400~kyr BP (before present) \cite{imbrie93}.  A similar ``Stage-1 problem'' occurred recently \cite{muller00}.  High-resolution spectral analysis of the glacial records shows a narrow 100~kyr cycle, in conflict with the double peak expected from the Milankovitch model \cite{muller97}.  Finally, the penultimate glacial termination appears to have preceded the insolation increase that is supposed to have caused it---a conflict known as the ``causality problem'' \cite{karner00,levine}.  Various \textit{ad hoc} solutions have been suggested for each of these problems on a case-by-case basis, but no common explanation has been found \cite{muller00b}.

In this paper we propose that the glacial cycles, rather than being driven by insolation cycles as in the present standard model, are initially driven by cosmic ray changes, probably through their effects on clouds.  There is a wide range of evidence that supports this model, although some of it is disputed.  The evidence, in chronological order, comprises:

\begin{enumerate}
  \item Svensmark \textit{et al.}\,\cite{svensmark9700} show a correlation between cloud cover and variations of cosmic ray flux.  These correlations, based on data from satellites and neutron monitors over the last two decades, have been subjected to intense scrutiny and criticism \cite{kernthaler}.  New analysis by Marsh and Svensmark \cite{marsh} support the original conclusion; however an analysis by Rohde \textit{et al.}\,\cite{rohde} shows that the effect may be more complicated than previously reported.  An increased flux of galactic cosmic rays (GCRs) is associated with increased low cloud cover, and therefore a cooler climate.  Mechanisms that could link cloud cover to cosmic radiation have been proposed, although none is experimentally established \cite{tinsley,yu,carslaw}.  Clouds are known to exert a strong effect on Earth's radiative energy balance (cloud cover reduces ground-level radiation by a global average of about 30 W/m$^2$, or 13\% of the ground-level solar irradiance), and so small changes would be climatically significant.  Furthermore a secular change of cloud cover would cause a net change in total ground-level insolation, whereas Milankovitch cycles result in no change of total insolation, but rather in a latitudinal and seasonal redistribution.  
  \item There is growing evidence for close links between climate changes in the Holocene (approximately the last 10~kyr) and variations of the cosmic ray flux (see, for example, \cite{bond97,bond01,neff}).  A high GCR flux is associated with a cold climate, and a low flux with a warm climate.  On these 100 yr timescales, variations of the cosmic ray flux are thought to reflect changing solar activity (increased solar magnetic activity reduces the GCR flux).  However, recent high-resolution paleomagnetic studies suggest that short-term geomagnetic variability may in fact control a significant fraction of the GCR modulation within the Holocene, even on 100 yr timescales \cite{stonge}.  Moreover, we will present evidence here that shows climate changes are also linked to variations of the cosmic ray flux caused by changes of geomagnetism.  This indicates that GCRs directly influence the climate, rather than merely serve as a proxy for solar variability.  We note, however, that the brief Laschamp event (37--40~kyr BP), when the geomagnetic field dropped to a low value, produced no evidence of climate change in the GRIP (Greenland) ice core \cite{wagner01}, although a pronounced reduction of the East Asia monsoon was registered in Hulu Cave, China \cite{wang}.
  \item Recent measurements suggest that long-term records of variations of Earth's magnetic field---in both strength and magnetic inclination---show the presence of orbital frequencies \cite{channell,yamazaki}.  The effects persist when the archives are corrected for climatic influences.  The 800~kyr record known as Sint-800 \cite{guyodo} is reported to show no such cycles; however we will present a new analysis here that shows such cycles are in fact present in this record.  It is perhaps surprising that orbital variations could modulate the Earth's dipole field, but we will argue that such a linkage is plausible.  In addition, we will discuss other mechanisms that could link orbital variations to changes of cosmic radiation.
  \item We present a new spectral analysis of the cosmic ray flux recorded in the \beten\ content of deep ocean sediments, which shows the presence of orbital cycles.  Furthermore we present additional results based on this \beten\ record and on precisely-dated speleothems that reinforce the causality problem with the Milankovitch model and support our conclusion that cosmic rays appear to be driving the glacial cycles.
\end{enumerate}

Bond \textit{et al.}\,\cite{bond97,bond01} have shown a strong correlation between ice-rafted-debris (IRD) events in the North Atlantic and increased GCR fluxes (measured both by \beten\ and \cft) during the last 12~kyr.  They assumed that the cause of the IRD events was reduced solar irradiance, and that the GCR fluxes simply provided a solar proxy.  The Little Ice Age around the 17th century appears to be the most recent of about 10 such centennial-scale events during the Holocene \cite{bond01}, when North Atlantic sea surface temperatures fell by around 2$^\circ$C in association with spells of anomalously high GCR flux (around +20\%).  Remarkably, the IRD events extend beyond the Holocene and into the last glaciation to at least 30~kyr BP, maintaining variability on a similar kilo-year timescale but with larger amplitude of debris \cite{bond97}.  This argues against a primary internal trigger such as ice sheet dynamics since the climate conditions varied a great deal over this interval.
 
Several previous studies have noted similarities between glacial climate records and variations of GCRs or paleomagnetism.  Sharma \cite{sharma} proposed that variations of solar activity control the glacial cycles.  Kok \cite{kok} interpreted the climate signal in the paleomagnetic data as evidence for inadequate correction of climatic effects on the archive deposition rates.  On much longer timescales, up to 1 Gyr, Shaviv \cite{shaviv02} reported evidence for the occurrence of ice-age epochs on Earth during crossings of the solar system with the galactic spiral-arms, when elevated GCR fluxes are expected.  This observation has recently been further supported by Shaviv and Veizer \cite{shaviv03}, who find a strong correlation between the GCR flux and reconstructed ocean temperatures during the Phanerozoic (past 550 Myr) \cite{veizer}.  Although these and other observations have been interpreted in diverse ways, we suggest that each is consistent with a single explanation, namely a direct effect on the climate by cosmic rays.

\section{COSMIC RAY RECORD AND TIMING OF THE GLACIAL CYCLES}
\label{sec_timing}
 
On reaching Earth, cosmic rays interact with nuclei in the atmosphere, creating showers of secondary particles, and dissipating energy by ionization.  Among the products are light radioisotopes, notably \cft\ and \beten, which settle on the surface of Earth either via the carbon cycle ($^{14}$CO$_2$) or in rain and snow (\beten).  Since this is their only terrestrial source, the resultant archives of light radioisotopes found in tree rings, ice cores and marine sediments provide records of the past GCR flux.  For the long-lifetime \beten\, these records extend back over several hundred thousand years.  To determine the GCR flux, each archive must first be corrected for transport effects.  In the case of \beten\ in deep-sea sediments, there are three main effects: (i) sediment redistribution and rate variation, (ii) boundary scavenging and (iii) variable ocean water circulation.  It has been shown that the effects of sediment redistribution and rate variation can be corrected by normalizing \beten\ to $^{230}$Th concentrations, yielding vertical settling rates into the sediments \cite{francois}.  Boundary scavenging and ocean water circulation effects are quantified and corrected by modeling the transport of \beten\ at the particular location of the archive \cite{anderson}. 
  
\begin{figure}[htbp]
  \begin{center}
      \makebox{\includegraphics[width=150mm]{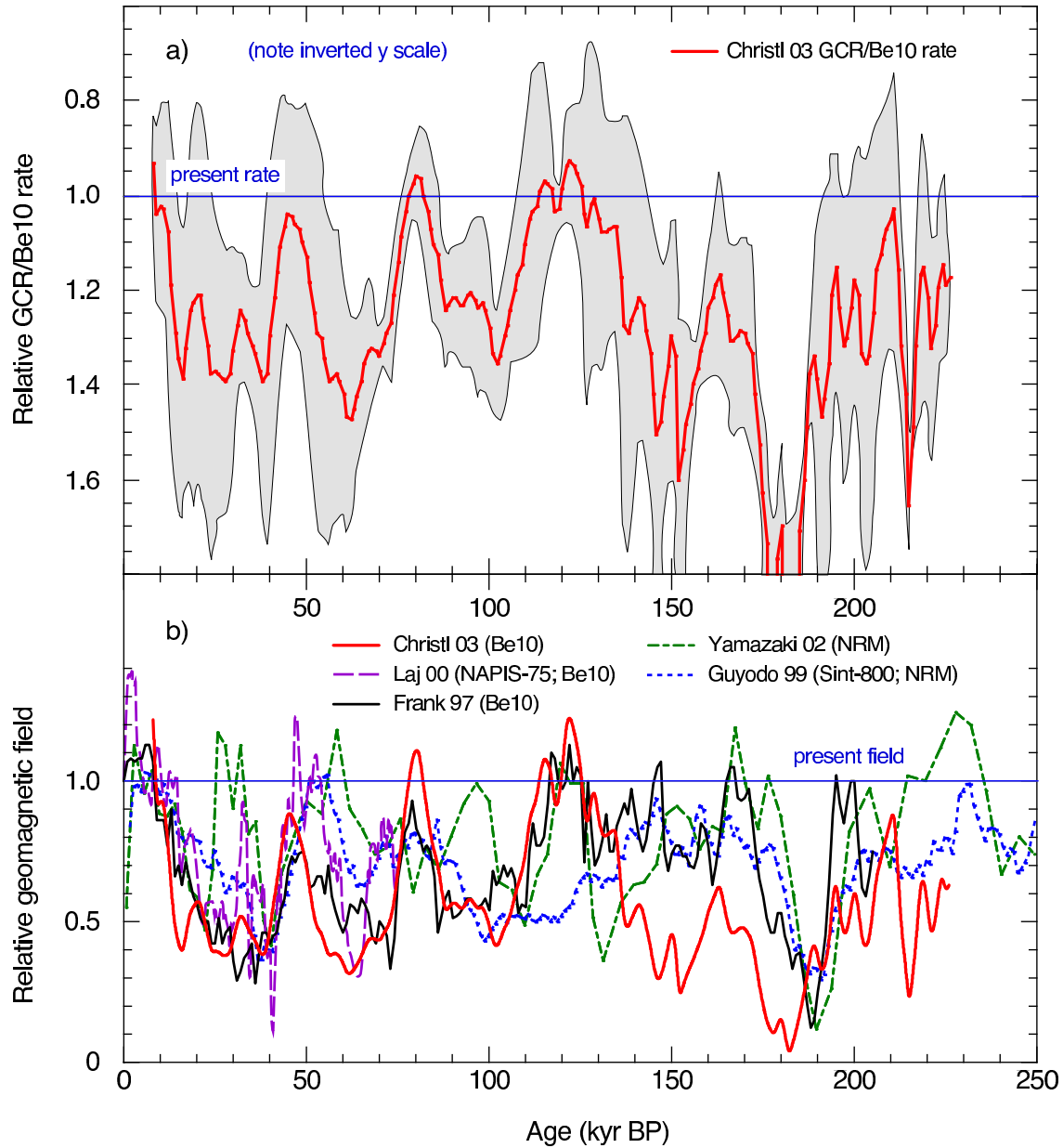}}
  \end{center}
  \caption{a) GCR intensity over the previous 220~kyr, reconstructed from combined \beten\ deep-sea sediments and based on the SPECMAP timescale \cite{christl}.  The grey shaded regions indicate the $1 \sigma$ confidence interval for the vertical scale.  The latter is inverted to follow the convention that warmer climates appear above cooler ones.  b) Relative geomagnetic field derived from the \beten\ data in panel a), assuming this to be the sole cause of the variations \cite{christl}.  Several other recent geomagnetic reconstructions \cite{yamazaki,guyodo,frank,laj} are shown for comparison.}  
  \label{fig_gcr_b_250kyr} 
  \end{figure}

Christl, Strobl, and Mangini \cite{christl} have recently reconstructed the global \beten\ production rate during the last 220~kyr using combined sediment cores from the Pacific, Atlantic and Southern Oceans (Fig.\,\ref{fig_gcr_b_250kyr}a).  The data were corrected for sediment redistribution and rate variation, and boundary scavenging---including glacial/interglacial variations.  The timescale is based on the SPECMAP stacked \doe\ record \cite{martinson}, which is tuned to precession and obliquity, and agrees with the 65$^\circ$N insolation cycles within an accuracy of about 5~kyr over this 200~kyr range.  An advantage of deep ocean sediment cores for the work presented here is the long residence time of \beten\ before settling (about 1~kyr).  This reduces sensitivity to regional variations of production or deposition, and ensures a measurement of global GCR flux.  In contrast, the atmospheric settling time of \beten\ into ice core archives is much shorter (between 1 wk  and 1 yr).
  
The \beten\ flux is modulated by both the geomagnetic field and the solar wind.  However, assuming for the moment that all variations on these long timescales are of geomagnetic origin, we can calculate the corresponding values of Earth's magnetic field \cite{wagner00}.  These are shown in Fig.\,\ref{fig_gcr_b_250kyr}b), along with several recent geomagnetic reconstructions for comparison, based on \beten\ in ocean sediments \cite{frank} and ice cores \cite{laj}, or paleointensity proxies derived from magnetic materials \cite{yamazaki,guyodo}.  To compensate for depositional variations due to climate, the paleointensity proxies are constructed by normalizing the natural remnant magnetization (NRM) by either the isothermal remnant magnetization (IRM) or the anhysteric remnant magnetization (ARM) \cite{tauxe}.
  
All data broadly agree that the geomagnetic field over the last 200~kyr was generally lower than at present, and displayed considerable variability on millennial timescales or greater.  However, there is only qualitative-to-poor agreement on the detailed structure.  In particular, discrepancies in the 110--190~kyr region within the NRM-based and the \beten-based reconstructions suggest either regional effects or experimental errors.  An additional potential difference between the NRM and \beten\ reconstructions is long-term solar variability, which directly affects only the \beten\ data.  For example, the increase of global GCR flux due to a low relative geomagnetic field of 0.6 could be cancelled by a high mean solar activity, approximately equal to present solar maxima (e.g.\,1991) \cite{wagner00}.
  
Our measurements of \beten\ production (Fig.~\ref{fig_gcr_b_250kyr}a) show that the GCR flux over the last 220~kyr was generally around 20--40\% higher than today, due to a reduced geomagnetic field.  However there were several relatively short periods when the GCR flux returned to present levels or, less frequently, increased further.  We have compared this GCR record with paleoclimatic data from stalagmites obtained in northern Oman and the central Alps.  The growth of stalagmites at each of these locations is especially sensitive to climate since it requires a climate at least as warm as today's.  In the case of Hoti Cave, Oman, water availability requires a warm climate and monsoon rains brought by a northward shift of the Inter Tropical Convergence Zone (ITCZ) in summer \cite{burns}.  Our studies of speleothems from Hoti Cave have shown that the ITCZ in Oman is strongly influenced during the Holocene by solar/GCR variability \cite{neff}.  The second cave, Spannagel, is situated at 2,500 m altitude in the Zillertal Alps, Austria, where the present mean air temperature is (1.5$\pm$1)\degc\ \cite{spotl02a}.  Stalagmites will not grow if the mean temperature falls below 0\degc\ because of permafrost.  The stalagmite samples are precisely dated by U/Th analysis to a $2 \sigma$ precision of 1~kyr (Spannagel) or 4~kyr (Hoti).
  
\begin{figure}[tbp]
  \begin{center}
      \makebox{\includegraphics[width=160mm]{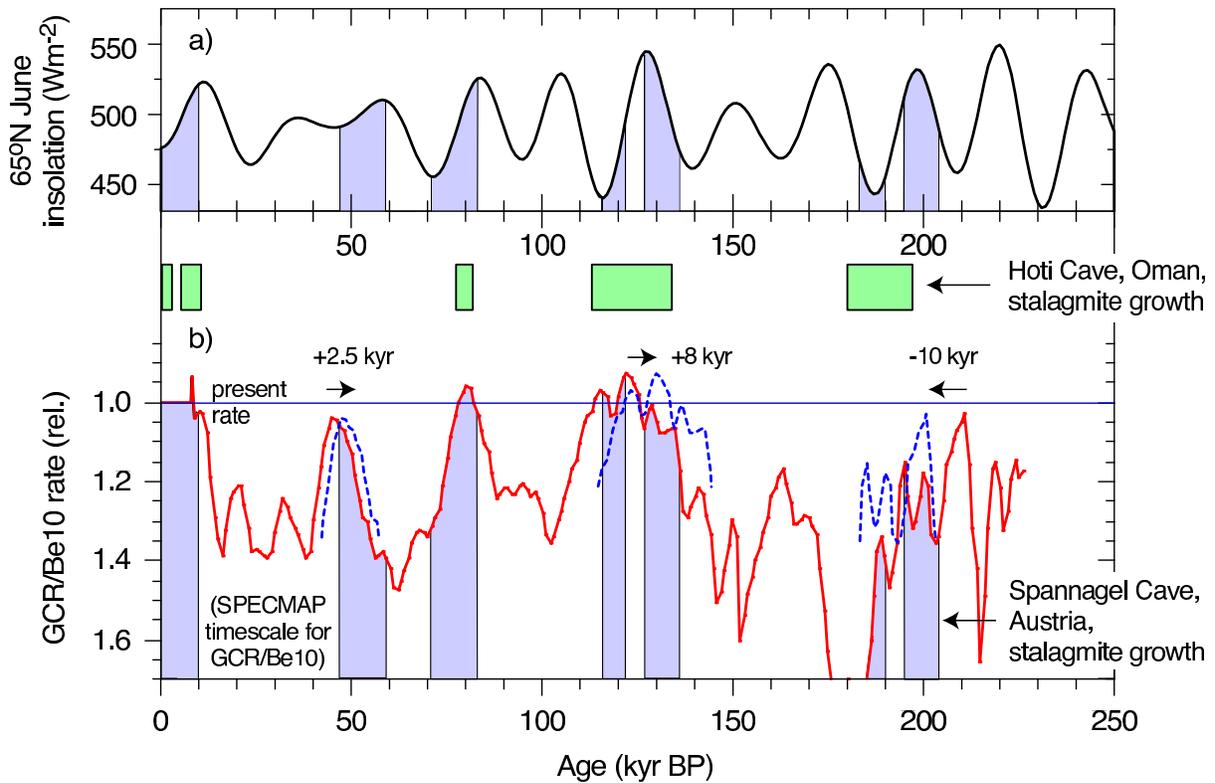}}
  \end{center}
  \caption{Comparison of the growth periods of stalagmites in Oman and Austria with a) 65$^\circ$N June insolation and b)~the relative GCR flux (\beten\ ocean sediments).  The growth periods are indicated by coloured regions, and require warm temperatures at these cave locations, close to the present climate. Periods without stalagmite growth are uncoloured.  The dashed curves in b) indicate the estimated corrections of systematic errors in the SPECMAP timescale, on which the GCR record is based \cite{spotl02a,henderson,spotl02b}.}  
  \label{fig_speleothem_growth} 
  \end{figure}

In Fig.\,\ref{fig_speleothem_growth} we compare the stalagmite growth periods with the GCR flux and the 65$^\circ$N June insolation.  We observe that most growth in Spannagel Cave occurs during intervals of low GCR flux (Fig.\,\ref{fig_speleothem_growth}b).  The growth periods are 0--10, 47--59, 71--83, 116--122, 127--136, 183--190 and 195--204~kyr BP.  In four of the five well-separated growth intervals in Spannagel Cave we also observe growth of stalagmites in Hoti Cave (0.2--2.8, 5.5--10.7, 78--82, 113--134 and 180--197~kyr BP).

In contrast, the stalagmite growth periods show no clear pattern of association with 65$^\circ$N June insolation (Fig.\,\ref{fig_speleothem_growth}a).  Growth would be expected to start around the time of insolation maxima, such as occurred for the final three periods starting at 10, 59 and 83~kyr.  However, at 136 and 204~kyr, the growth begins near the insolation minima---and the earlier growth period is interrupted and then re-commences at 190~kyr, also near an insolation minimum.  It is therefore difficult to attribute the warm growth periods at these two sensitive locations to northern summer insolation forcing.
  
\begin{figure}[htbp]
  \begin{center}
      \makebox{\includegraphics[width=160mm]{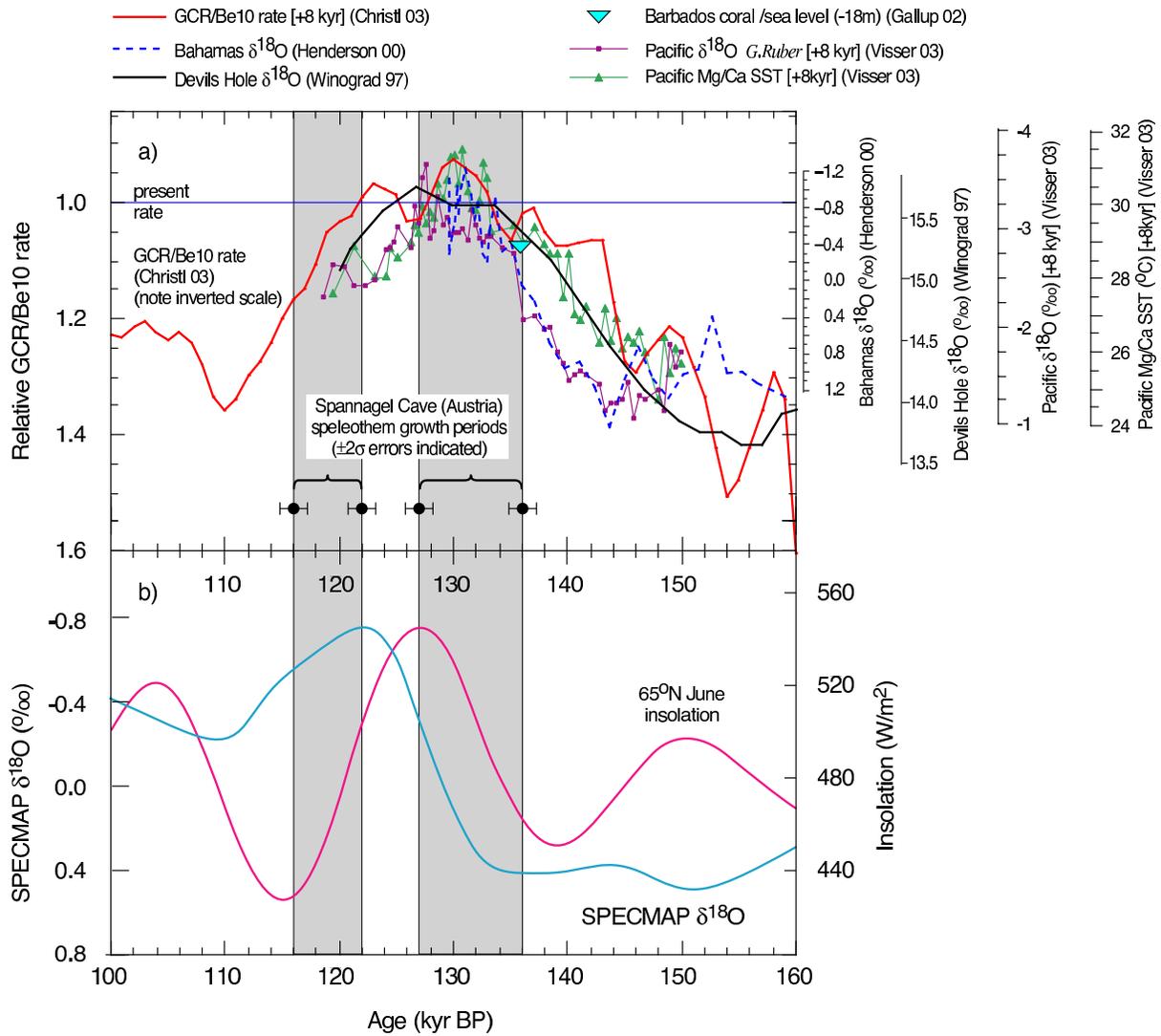}}
  \end{center}
  \caption{Timing of glacial Termination II.  a) The GCR rate together with the Bahamian \doe\ record \cite{henderson}, the date when the Barbados sea level was within 18 m of its present value \cite{gallup}, the \doe\ temperature record from Devils Hole cave, Nevada \cite{winograd}, and the Visser \textit{et al.}\,measurements \cite{visser} of the Indo-Pacific Ocean surface temperature and \doe\ records.  The GCR rate and the Visser \textit{et al.}\,data are shifted earlier by 8kyr in order to correct for estimated systematic errors in the SPECMAP timescale, on which they are based.  The growth periods of stalagmite SPA 52 from Spannagel Cave, Austria, are indicated by grey bands and black points.  b) The 65$^\circ$N June insolation and the SPECMAP \doe\ record \cite{martinson}.}  
  \label{fig_termination2} 
  \end{figure}

In Fig.\,\ref{fig_termination2}, we examine the timing of Termination II in more detail.  The \beten\ data show that the GCR flux began to decrease around 150~kyr, and it had reached present levels by about 135~kyr.  The growth period of stalagmite SPA 52 from Spannagel Cave began at 135$\pm$1.2~kyr.  So, by that time, temperatures in central Europe were within (1.5$\pm$1)\degc\  of the present day.  This corroborates Henderson and Slowey's conclusion, based on sediment cores off the Bahamas, that warming was well underway at 135$\pm$2.5~kyr \cite{henderson}.  Furthermore, dating of a Barbados coral terrace shows that the sea level had risen to within 20\% of its peak value by 135.8$\pm$0.8~kyr \cite{gallup}.  These results confirm the ``early'' timing of Termination II, originally discovered at Devils Hole Cave, Nevada \cite{winograd}.  We conclude that the warming at the end of the penultimate ice age was underway at the minimum of 65$^\circ$N June insolation, and essentially complete about 8~kyr prior to the insolation maximum.
  
Based on an analysis of deep ocean cores, Visser \textit{et al.}\,\cite{visser} have recently reported that warming of the tropical Pacific Ocean during Termination II preceded the northern ice sheet melting by 2--3~kyr.  Both their data and our GCR record are adjusted to the SPECMAP \doe\ timescale \cite{martinson} which is in turn tuned to the insolation cycles.  Relaxing this constraint, we have adjusted the timescales (earlier by 8~kyr) of both the Visser \textit{et al.}\,data and our GCR record to agree with the precisely- and independently-dated \doe\ data of Henderson and Slowey \cite{henderson}.  The prescient warming of the tropical sea surface temperature dramatically underscores the causality problem facing the Milankovitch model.  All data, however, are compatible with Termination II being driven by a reduction of cosmic ray flux.  We note that a similar reduction of GCR flux also occurred at Termination I due to a rise in geomagnetic field strength that started around 20~kyr BP \cite{bard}, coincident with the first signs of warming at the end of the last glaciation \cite{lao}.

\section{SPECTRAL ANALYSES OF GEOMAGNETISM AND COSMIC RAY FLUX}
\label{sec_spectral_analysis}

There is a long history of papers reporting orbital signals in records of Earth's magnetism, and an equally long history of papers disputing those claims.  In this section, we'll review only the recent history from 1998 until the present.  All of these papers applied the Blackman-Tukey method \cite{blackman}, mostly using the AnalySeries software package \cite{paillard}. Unfortunately, none reported the key ``lag'' parameter of this program, so we assume the default value of 1/3 was used, whereas it should be varied between 0 and 1 in order to find the optimum value of resolution while still compensating for uncertainties in timescale.  We have therefore repeated the spectral analyses of each of these papers to confirm the original conclusions \cite{power}.
   
\begin{figure}[htbp]
  \begin{center}
      \makebox{\includegraphics[width=117mm]{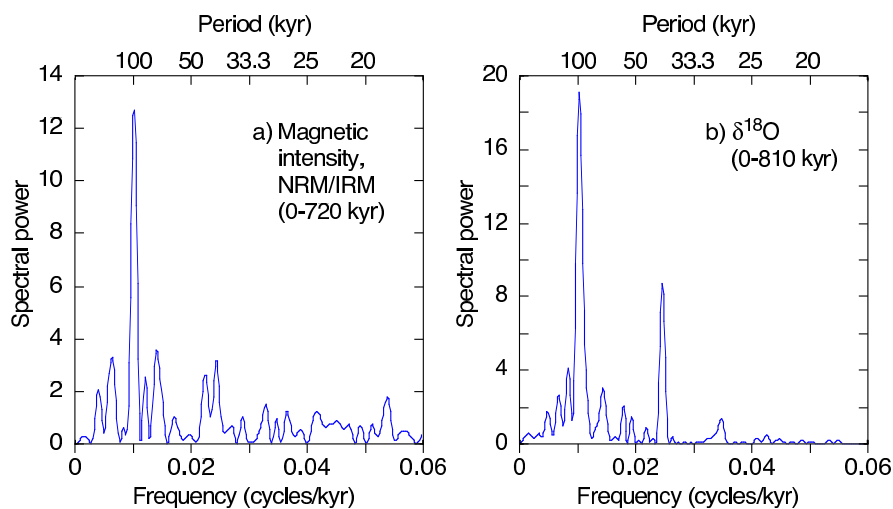}}
  \end{center}
  \caption{Spectral powers of a) the geomagnetic paleointensity (mean NRM/IRM) of site 983 \cite{channell}, and b) the \doe\ in the benthic stack of Karner \textit{et al.}\,\cite{karner02}.}  
  \label{fig_nrm_o18} 
  \end{figure}

In 1998, Channell \textit{et al.}\,\cite{channell} measured the geomagnetic paleointensity from two sea-floor cores, ODP Sites 983 (to 720 ka) and 984 (to 430 ka).  Their spectral analyses showed two strong spectral peaks corresponding to 100~kyr and 41~kyr, which match orbital and climate periodicities. To be cautious, they claimed unambiguous discovery only of the 41~kyr cycle.  Our reanalysis of their data confirms the statistical significance of a 100~kyr peak.  In Fig.\,\ref{fig_nrm_o18} we show our estimate of the spectral power for the paleointensity (mean NRM/IRM) of site 983.  There is a strong peak at 0.01 c/kyr (period 100~kyr), and power near 0.024 c/kyr (41~kyr) that is statistically significant at the 95\% CL.  For comparison, we plot the spectral power for \doe\ in the benthic stack of Karner \textit{et al.}\,\cite{karner02}.  The correlation they report between the 100~kyr cycle in paleointensity and in IRM shows that some of the 100~kyr signal may be spurious.  This is a cautious conclusion; a large part of the 100~kyr signal may still be magnetic.  The presence of the 100 and 41~kyr periods is robust.  We have reanalyzed the data with a series of windows (Bartlett, Hanning, Welch, Chebyshev, and others), and with the Maximum Entropy method; the strength of the two lines persists. Furthermore, when the data are split into two consecutive time periods, the two peaks are present in both halves. 
   
\begin{figure}[htbp]
  \begin{center}
      \makebox{\includegraphics[width=150mm]{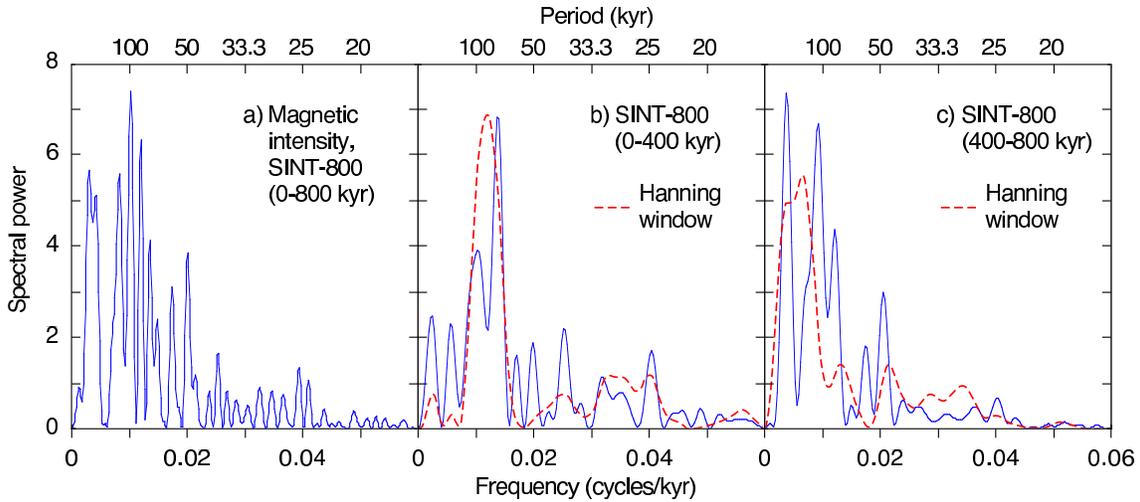}}
  \end{center}
  \caption{Spectral powers of the Sint-800 paleointensity record \cite{guyodo}: a) 0--800~kyr, b) 0--400~kyr and c) 400--800~kyr.  The dashed lines are the spectra when a Hanning filter is applied to the data; such a filter broadens peaks by a factor of about 1.5, but smooths unphysical fluctuations that arise from the abrupt termination of the records.}  
  \label{fig_sint800} 
  \end{figure}

In 1999, Guyodo and Valet \cite{guyodo} published a stack of 33 globally-distributed paleointensity records, named Sint-800 (since it goes back 800~kyr).  They concluded from their spectral analysis that Sint-800 does not indicate the presence of any dominant periodicity \cite{lag}.  In Fig.\,\ref{fig_sint800}a) we show a periodogram of Sint-800.  The rapid variations in the low frequency region, 0--0.02 c/kyr, suggest an inaccurate timescale.  A 5\% error in the Sint-800 data at 800~kyr (from Fig.\,2 in \cite{guyodo}) amounts to 40~kyr---enough to cancel not only a 41~kyr cycle but also a 100~kyr cycle.  In Figs.\,\ref{fig_sint800}b) and c) we therefore show periodograms of Sint-800 separately for 0--400~kyr and 400--800~kyr, with both flat and Hanning windows \cite{muller00}.  (A similar analysis was made in \cite{guyodo}, but with lower resolution than either of our methods.)  The 0--400~kyr period shows a strong peak at 0.01 c/kyr.  There is a smaller peak near 0.024 c/kyr.  The 400--800~kyr period shows no significant structure; the power below 0.02 c/kyr is similar to the `red' (low frequency) noise that one expects when no strong periodicity is present---or when there are timescale errors.  In summary, our analysis shows that a strong 100~kyr cycle is present in the first half of Sint-800; the 41~kyr cycle is more sensitive to timescale errors and may have cancelled out.  We conclude that the absence of Earth's orbital frequencies in the full Sint-800 record could be due to an inaccurate timescale.
   
\vspace{5mm}   
   
\begin{figure}[htbp]
  \setlength{\unitlength}{1mm}
 \hfill
\begin{minipage}[b]{65mm}  
\noindent
  \begin{center}
      \makebox{\epsfig{file=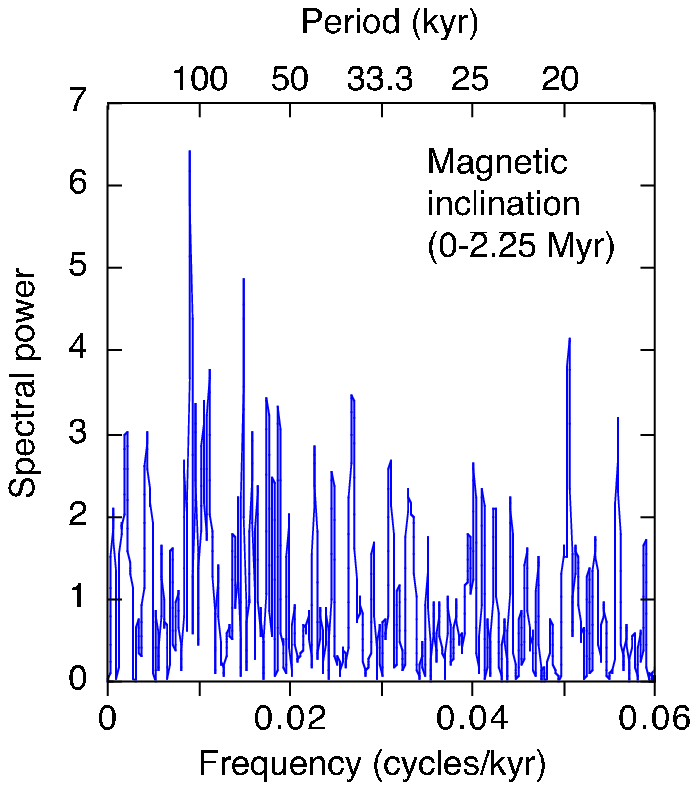,width=60mm}}
  \end{center}
  \caption{Spectral power of Earth's magnetic inclination, 0--2.25 Myr \cite{yamazaki}.}  
  \label{fig_inclination} 
\end{minipage}
\hfill
\begin{minipage}[b]{65mm}  
\noindent
  \begin{center}
      \makebox{\epsfig{file=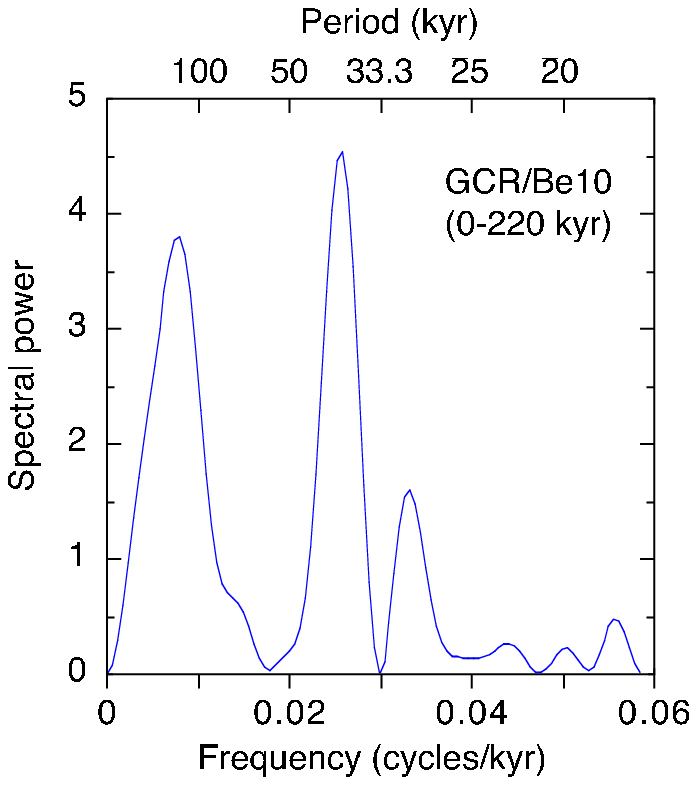,width=60mm}}
  \end{center}
  \caption{Spectral power of the GCR flux (\beten\ ocean sediments), 0--220~kyr.}  
  \label{fig_be10} 
\end{minipage}
 \hfill \mbox{}
\end{figure}

In 2002, Yamazaki and Oda presented measurements of the intensity and direction of Earth's magnetic field back to 2.25 Myr BP, and reported discovery of a 100~kyr periodicity in inclination \cite{yamazaki}.  A measurement of magnetic inclination has the advantage that it does not require compensation for sedimentation rate; all it needs is a good timescale.  In Fig.\,\ref{fig_inclination} we show a periodogram of their inclination record for 0--2.25 Myr.  The highest peak is near 0.01 c/kyr, with a height of 6.6.  The mean in the region 0--0.2 c/kyr is 1.1; thus the peak is a factor of 6 above background.  Since there are about 60 independent frequencies in the plot, the probability of a random peak of this magnitude is 60 $e^{-6}$ = 0.15, yielding significance at 85\% CL.  However, since the 100~kyr period is of interest \textit{a priori}, one can ask if there is significant power in this region alone (within 10\% of the 100~kyr period).  Since there are only 2 independent frequencies in this region, the probability of such a random fluctuation becomes 2 $e^{-6}$ = 0.005, making the peak significant at 99.5\% CL.  Thus, we agree with the conclusion of Yamazaki and Oda that such a period is present.  However, we note that the area under the 100~kyr region is small; the 100~kyr period accounts for only 8\% of the variance. 
   
In Fig.\,\ref{fig_be10} we show a periodogram of the GCR flux during the last 220~kyr (from Fig.\,\ref{fig_gcr_b_250kyr}a).  Although the resolution is poor (wide peaks) because of the short record, it is nevertheless adequate to test for the presence of orbital cycles.  The two strongest peaks are consistent with the frequencies reported in the geomagnetic records, namely 0.01 c/kyr (100~kyr period) and 0.024 c/kyr (41~kyr).  The frequency match is not perfect, but neither should it be; Monte Carlo simulations show that the frequency estimate has a standard error approximately equal to the half-width at half-maximum \cite{muller00}.  We find the strengths of these two peaks are insensitive to possible systematic effects resulting from the reduced time range (10--160~kyr) of one of the 4 records that comprise the \beten\ stack \cite{christl}.
    
We conclude that there is suggestive evidence that Earth's magnetic field and the GCR flux have components that vary with orbital frequencies.  Various methods have been used to reduce the danger that climate variations could simulate changes of the Earth's field or of the \beten\ flux, e.g.\,through variations in the concentration of remanence-carrying grains or variations in sedimentation rate.  (We note, moreover, that sedimentation rates determined by orbital tuning have never shown 100 or 41~kyr periods \cite{muller00}.  Furthermore, at least one of the proxies---magnetic inclination---is unaffected by these systematic problems.)  Even though the presence of such signals is not definitive, we suggest that previous conclusions that orbital frequencies are absent were premature.  The most robust signal is the 100~kyr cycle, although there is some evidence for a 41~kyr cycle.  Inaccurate timescales tend to hide such periodicity; a 20~kyr error could cancel a 41~kyr period, and a 50~kyr error could cancel the 100~kyr period.  The 100~kyr period could be due to either orbital eccentricity or orbital inclination; the 41~kyr period could be due to obliquity (the tilt of Earth's pole).  The presence of such frequencies was not predicted and the cause of the linkage is not obvious.  We will discuss several possible explanations in the next section.

\section{COSMIC RAY MODULATION}
\label{sec_gcr_modulation}

The GCR flux incident on Earth's atmosphere is modulated by three processes: a) variations of the solar wind within the heliosphere (on 10--1000 yr timescales, and possibly longer), b) variations of Earth's magnetic field (100--10,000 yr), and c) variations of the interstellar flux outside the heliosphere \mbox{$(>$10 Myr).}  On entering the heliosphere, GCRs sense the heliospheric magnetic field (HMF) of the solar wind, about which they spiral \cite{parker}.  To reach Earth they must overcome the effects of outward convection and irregularities in the HMF, which impose a random walk on the particle motion.  In addition to convection and diffusion, the GCRs undergo coherent drift motion in the large-scale ordered structure of the HMF.  On reaching Earth, cosmic rays must traverse the geomagnetic field to reach the lower atmosphere.  In consequence, the GCR intensity is about a factor 4 higher at the poles than at the equator, and there is a more marked solar cycle variation at higher latitudes.  
    
The GCR flux over these different timescales varies by between 15\% during the 11 yr solar cycle, to as much as a factor 2 increase during periods of low geomagnetic field and low solar activity.   Interstellar modulations of the GCR flux are estimated to be between -75\% and +35\% of present values \cite{shaviv02} on cosmological timescales, corresponding to the 140 Myr crossing period of the solar system with the spiral arms of the Milky Way (where the peak fluxes probably reside).  Nearby supernovae could increase the GCR fluxes above these values.   In summary, if the cosmic ray-climate connection is causal, then the climate appears to be remarkably sensitive to quite small secular changes of GCR intensity---of around 10\% or so.
    
There are two requirements that underpin the proposed GCR model for the glacial cycles: 1) the existence of a plausible mechanism by which GCRs exert a significant effect on the climate, and 2) the existence of a plausible mechanism by which the GCR flux is orbitally modulated.  At this stage, neither requirement need be proved---nor indeed can be.  It is nevertheless of value to propose a model based on physics since it leads to predictions that can be tested. 
     
The first requirement is satisfied by the cosmic ray-cloud interaction suggested by satellite observations \cite{svensmark9700}.  If a causal mechanism is confirmed, this could provide an effective initial step by which an energetically-weak GCR signal (which is roughly equivalent to that of starlight) is amplified into a significant climate forcing.  Physical mechanisms connecting GCRs with clouds and thunderstorms have been proposed \cite{carslaw,iaci_ws}, including: a) ion-induced cloud condensation nuclei (CCN) production \cite{yu,eichkorn}, b) electrofreezing of supercooled liquid droplets \cite{tinsley}, c) direct effects on thunderstorm electrical processes, and d) influences on the global electrical circuit.  However most of the associated microphysical processes have yet to be investigated experimentally and quantified. 
      
Mechanisms for the second requirement are more speculative.  Although we are unable to account for a 100~kyr cycle in the GCR flux, we can imagine two potential mechanisms: 1) orbital modulation of the geodynamo and hence of the geomagnetic field strength and direction, and 2) correlations of orbital parameters with the solar wind shielding in the heliosphere. 
      
A physical mechanism that could in principle link orbital gravitational forces to the geodynamo is a difference in the precessional torques on Earth's mantle and its solid iron core, which are separated by a liquid iron outer core \cite{malkus}.  It is the motion of the liquid core that is thought to be responsible for the geodynamo, so any relative motion induced between the mantle and inner core would affect the geomagnetic field.  Orbital modulation of the geodynamo has some experimental support, as described above.  We remark that a GCR-cloud-climate effect may depend not only on the globally-averaged GCR flux but, in particular, on its variation at low latitudes.  In this case, variations in declination must also be considered since they could produce significant latitudinal variations without any change in overall geomagnetic field strength.  We also note that low latitudes are where geomagnetic variations most strongly affect the GCR fluxes (in contrast with solar modulation, which is more important at high latitudes).
      
The second mechanism would require the GCR flux to have a significant heliographic latitudinal dependence.  Muller and MacDonald \cite{muller97} have shown that the inclination of Earth's orbital plane varies with a 100~kyr period by up to 3 degrees relative to its average position.   If the average GCR flux had a significant heliographic latitudinal dependence then this could introduce a 100~kyr modulation component.  Cosmic ray modulation in the heliosphere is characterized by a dominance of transient disturbances during solar maximum and, during solar minimum, of large-scale drifts in the HMF and wavy current sheet \cite{issi98}.  Measurements by Ulysses \cite{issi98} out of the ecliptic have shown latitudinal gradients consistent with zero at solar maximum, and a small but finite gradient at solar minimum of about +0.2\% /degree (and up to +1\% /degree close to the ecliptic) \cite{hmf}.  To be significant, a stronger latitudinal gradient would be required, implying a magnetically-quiet Sun and a thin equatorial heliospheric current sheet.  There is some indirect evidence that this could occur.  The existence of the Maunder Minimum (and its ten cousins earlier in the Holocene) has demonstrated that the Sun can enter a quiet state for about 100 years characterized by periodic polarity reversals but an almost complete absence of magnetic disturbances (few sunspots or aurorae).  These conditions produce higher GCR fluxes and may also give rise to increased latitudinal gradients.  The open question is whether the Sun could remain in such a state for longer periods, which may be sufficient to trigger a glaciation in association with a ``favourable'' orbital enhancement of the GCR flux.

\section{CONCLUSIONS}
\label{sec_conclusions}

We are proposing a new model for the glacial cycles in which the forcing mechanism is due to galactic cosmic rays, probably through their effect on clouds.  We are led to this conclusion from the accumulated experimental evidence of the last few years as well as new results presented here on a 220~kyr record of GCR flux obtained in deep-ocean sediments.  Although the evidence is sufficient to propose the GCR model for the glacial cycles, it is clearly insufficient to establish it.  Nevertheless the model makes definite predictions that can be tested by further observations and experiments. 
       
The first area to be tested concerns the paleo record of GCR flux, its orbital components and association with climate change.  Further \beten\ measurements in sediment cores are required, over longer time spans and with improved precision and dating.  Parallel improvements are required for paleomagnetic intensity and direction in order to study the orbital components and to separate solar and geomagnetic effects in the \beten\ record.  Orbital influences on the geomagnetic field should be modeled.  Further satellite data on GCR/solar wind characteristics in the heliosphere are required, both in and out of the ecliptic, and during different periods of solar magnetic activity.  GCR transport in the heliosphere for a magnetically quiet Sun (Maunder Minimum) should be modeled to estimate the expected magnitude of orbital variations of the GCR flux.
       
The second area to be tested concerns the interactions of GCRs with Earth's clouds and climate.  Improved and extended satellite observations of clouds are needed.  Investigations are required on the effects of GCRs on clouds and thunderstorms, including ion-induced CCN production, electrofreezing of supercooled liquid droplets and atmospheric electrical processes.  The microphysics of GCR-cloud-climate interactions should be investigated in laboratory experiments under controlled conditions, and the results applied to models and field observations.  Combined interdisciplinary efforts in these directions may quite quickly be able to establish whether or not the GCR model for the glacial cycles is further supported, and where more work is needed to quantify its physical basis.

\section*{Acknowledgements}

We would to like to thank M. Christl, D. Karner and R.A. Rohde for valuable critiques of this paper.


\begin{thebibliography}{99}

\bibitem{do18} The physical basis for proxy climate measurements from the stable $^{18}$O isotope is that the vapour pressure of \htoe\ is lower than that of \htos. Evaporation from the oceans thus produces water vapour that is $^{18}$O-depleted (by about 1\% relative). The relative proportion of $^{18}$O in a sample is expressed in terms of its fractional deviation, \doe, from a standard mean ocean water value (about $2 \cdot  10^{-3}$).  The \doe\ value of sea sediments provides a measure  of the global volume of water locked up in ($^{18}$O-depleted) ice sheets, since high ice volumes leave the oceans enriched in $^{18}$O. 
 
\bibitem{imbrie80} A nonlinear ice model that accomplishes this was first proposed by J. Imbrie and J.Z. Imbrie, \textit{Science} \textbf{207}, 943 (1980).
 
\bibitem{muller00} R.A. Muller, G.J. MacDonald, \textit{Ice Ages and Astronomical Causes}, Springer Praxis, Chichester, UK, (2000). 

\bibitem{muller00a} Muller and MacDonald \cite{muller00} showed that virtually any nonlinear operation on insolation has the same effect. 

\bibitem{imbrie93} J. Imbrie \textit{et al.}, \textit{Paleoceanography} \textbf{8}, 699 (1993).

\bibitem{muller97} R.A. Muller, G.J. MacDonald, \textit{Science} \textbf{277}, 215 (1997).

\bibitem{karner00} D.B. Karner, R.A. Muller, \textit{Science} \textbf{288}, 2134 (2000).

\bibitem{levine} For a recent review, see J. Levine, D.B. Karner, R.A. Muller, \textit{Eos} \textbf{82} (47) (Fall Meeting Suppl.) abstr. U12A-0004 (2001).

\bibitem{muller00b} Muller and MacDonald \cite{muller00} proposed that variations in the accretion of extraterrestrial dust could explain all the data, but the amount of extraterrestrial dust accreting on Earth's atmosphere is less than 1 $\mu$m per year, and it has been questioned whether this could drive cloud cover.

\bibitem{svensmark9700} H. Svensmark, E. Friis-Christensen, \textit{J.\,Atmos.\,Solar\,Terr.\,Phys.} \textbf{59}, 1225 (1997); N. Marsh, H.~Svensmark, \textit{Phys. Rev. Lett.} \textbf{85}, 5004, (2000). 

\bibitem{kernthaler} S. C. Kernthaler, R. Toumi, J.D. Haigh, \textit{Geophys.\,Res.\,Lett.} \textbf{26}, 863 (1999); T. B. Jorgensen, A.W. Hansen, \textit{J.\,Atmos.\,Solar\,Terr.\,Phys.} \textbf{62}, 73 (2000); J.E. Kristj‡nsson, J. Kristiansen, \textit{J.\,Geophys.\,Res.} \textbf{105}, 11851 (2000); J.E. Kristj‡nsson, A. Staple, J. Kristiansen, \textit{Geophys.\,Res.\,Lett.} \textbf{29}, 10.1029/2002GL015646 (2002); B. Sun, R.S. Bradley, \textit{J.\,Geophys.\,Res.} \textbf{107}, D14, 10.1029/2001JD000560 (2002).

\bibitem{marsh} N. Marsh, H. Svensmark, \textit{J.\,Geophys.\,Res.} \textbf{108}, D6, 10.1029/2001JD001264 (2003).

\bibitem{rohde} R.A. Rohde, J. Levine, R.A. Muller, \textit{Eos}. Trans. AGU 83 (47), Fall Meet. Suppl., Abstract GC21-B-0167, p. F397 (2002).

\bibitem{tinsley} B.A. Tinsley, G.W. Deen, \textit{J.\,Geophys.\,Res.} \textbf{96}, 22283 (1991); B.A. Tinsley, \textit{J.\,Geophys.\,Res.} \textbf{101}, 29701 (1996); B.A. Tinsley, \textit{Space Sci.\,Rev.} \textbf{94}, 231 (2000).

\bibitem{yu}  F. Yu, R.P. Turco, \textit{J.\,Geophys.\,Res.} \textbf{106}, 4797 (2001); F. Yu, \textit{J.\,Geophys.\,Res.} \textbf{107}, A7, 10.1029/2001JA0D0248 (2002); L. Laakso \textit{et al.}, \textit{J.\,Geophys.\,Res.} 10.1029/2002JD002140 (2002).

\bibitem{carslaw} K.S. Carslaw, R.G. Harrison, J. Kirkby, \textit{Science} \textbf{298}, 1732 (2002).

\bibitem{bond97} G.C. Bond \textit{et al.}, \textit{Science} \textbf{278}, 1257 (1997). 

\bibitem{bond01} G.C. Bond \textit{et al.}, \textit{Science} \textbf{294}, 2130 (2001).
 
\bibitem{neff} U. Neff, S.J. Burns, A. Mangini, M. Mudelsee, D. Fleitmann, A. Matter, \textit{Nature} \textbf{411}, 290 (2001).
 
\bibitem{stonge} G. St-Onge, J.S. Stoner, C. Hillaire-Marcel, \textit{Earth and Planet.\,Sci.\,Lett.} \textbf{209}, 113 (2003).
 
\bibitem{wagner01} G. Wagner \textit{et al.}, \textit{J.\,Geophys.\,Res.}, \textbf{106}, D4, 3381 (2001).
 
\bibitem{wang} Y.J. Wang \textit{et al.}, \textit{Science} \textbf{294}, 2345 (2001).
 
\bibitem{channell} J.E.T. Channell \textit{et al.}, \textit{Nature} \textbf{394}, 464 (1998).
 
\bibitem{yamazaki}  T. Yamazaki, H. Oda, \textit{Science} \textbf{295}, 2435 (2002).
 
\bibitem{guyodo} Y. Guyodo, J.-P.Valet, \textit{Nature} \textbf{399}, 249 (1999).
 
\bibitem{sharma} M. Sharma, \textit{Earth and Planet.\,Sci.\,Lett.} \textbf{199}, 459 (2002).
 
\bibitem{kok} Y.S. Kok, \textit{Earth and Planet.\,Sci.\,Lett.} \textbf{166}, 105 (1999).
 
\bibitem{shaviv02} N.J. Shaviv, \textit{Phys. Rev. Lett.} 89.051102 (2002).
 
\bibitem{shaviv03} N.J. Shaviv, J. Veizer, \textit{GSA Today}, Geological Society of America, \textbf{4} (July 2003).
 
\bibitem{veizer} J. Veizer, Y, Godderis, L.M. Franois, \textit{Nature} \textbf{408}, 698 (2000).
 
\bibitem{francois} R. Francois, M. P. Bacon \textit{et al.}, \textit{Paleoceanography} \textbf{5}, 761 (1990); M. Frank, R. Gersonde \textit{et al.}, \textit{Geol.\,Rundschau} \textbf{85}, 554 (1996); B. Schwarz, A. Mangini \textit{et al.}, \textit{Geol.\,Rundschau} \textbf{85}, 536 (1996).
 
\bibitem{anderson} R.F. Anderson \textit{et al.}, \textit{Earth and Planet.\,Sci.\,Lett.} \textbf{96}, 287 (1990); Y. Lao \textit{et al.}, \textit{Earth and Planet. Sci.\,Lett.} \textbf{113}, 173 (1992).
 
\bibitem{christl} M. Christl, C. Strobl, A. Mangini, \textit{Quat. Sc. Rev.} \textbf{22}, 725 (2003).
 
\bibitem{martinson} D. G. Martinson, N.G. Pisias, J.D. Hays, J. Imbrie, T.C. Moore, N.J. Shackleton, \textit{Quat.\,Res.} \textbf{27}, 1 (1987).
 
\bibitem{wagner00} G. Wagner \textit{et al.}, \textit{Nucl.\,Inst.\,Meth.} \textbf{B172}, 597 (2000).
 
\bibitem{frank} M. Frank \textit{et al.}, \textit{Earth and Planet.\,Sci.\,Lett.} \textbf{149}, 121 (1997).
 
\bibitem{laj} C. Laj, C. Kissel, A. Mazuad, J.E,T Channell, J. Beer, \textit{Phil.\,Trans.\,R.\,Soc.\,Lond.} \textbf{A 358}, 1009 (2000).
 
\bibitem{tauxe} L. Tauxe, \textit{Rev.\,of Geophysics} \textbf{31}, 319 (1993).
 
\bibitem{burns} S.J. Burns, D. Fleitmann, A. Matter, U. Neff, A. Mangini, \textit{Geology } \textbf{29}, 7, 623 (2001).
 
\bibitem{spotl02a} C. Sp\"{o}tl, A. Mangini, N. Frank, R. Eichst\"{a}dter, S.J. Burns, \textit{Geology } \textbf{30}, 9, 815 (2002).  
 
\bibitem{henderson} G.M. Henderson, N.C. Slowey, \textit{Nature} \textbf{404}, 61 (2000).
 
\bibitem{gallup} C.D. Gallup, H. Cheng, F.W. Taylor, R.L. Edwards, \textit{Science} \textbf{295}, 310 (2002).
 
\bibitem{winograd} I.J. Winograd, T.B. Coplen \textit{et al.}, \textit{Science} \textbf{258}, 255 (1992).
 
\bibitem{visser} K. Visser, R. Thunell, L. Stott, \textit{Nature} \textbf{421}, 152 (2003).
 
\bibitem{bard} E. Bard, B. Hamelin, R.G. Fairbanks, A. Zindler, \textit{Nature} \textbf{345}, 405 (1990).
 
\bibitem{lao} Y. Lao, R.F. Anderson \textit{et al.}, \textit{Nature} \textbf{357}, 576 (1992).
 
\bibitem{blackman} R.B. Blackman, J.W. Tukey, \textit{The Measurement of Power Spectra from the Point of View of Communication Engineering}, Dover, New York (1958).
 
\bibitem{paillard} D. Paillard, L. Labeyrie, P. Yiou, \textit{Eos} \textbf{77}, 379 (1996).
 
\bibitem{power} The spectral power was calculated by interpolating the data to equally spaced points, removing the average (but not the trend), using a boxcar window (i.e.\,all points equally weighted), padding the end with zeroes (to give intermediate frequencies), and then taking the square of the Fourier transform.
 
\bibitem{karner02} D.B. Karner, J. Levine, B.P. Medeiros, R.A. Muller, \textit{Paleoceanography} \textbf{17}, 10.1029/2001PA000667 (2002).
 
\bibitem{lag} This spectral analysis may have been hampered by use of the low resolution value of 1/3 for the Blackman-Tukey lag parameter.
 
\bibitem{parker} E.N. Parker, \textit{Planetary Space Sci.} \textbf{13}, 9 (1965).
 
\bibitem{iaci_ws} \textit{Proc. of the Workshop on Ion-Aerosol-Cloud Interactions}, ed. J. Kirkby, CERN, Geneva, CERN 2001-007 (2001).
 
\bibitem{eichkorn} S.Eichkorn \textit{et al.}, \textit{Geophys.\,Res.\,Lett.} \textbf{29}, 43-1 (2002).
 
\bibitem{malkus} W.V.R. Malkus, \textit{Science} 160, 259 (1968).
 
\bibitem{issi98} \textit{Cosmic Rays in the Heliosphere}, eds. LA. Fisk, J.R. Jokipii, G.M. Simnett, R. von Steiger, K.-P. Wenzel, Space Science Series of ISSI, Kluwer Academic Publishers, Dordrecht (1998).
 
\bibitem{hmf} The sign of the heliographic latitudinal gradients may reverse for opposite polarity of the HMF.
 
\bibitem{spotl02b} C. Sp\"{o}tl, A. Mangini, \textit{Earth and Planet.\,Sci.\,Lett.} \textbf{203}, 507 (2002).

 
\end{thebibliography}
\end{document}